\documentclass[12pt]{iopart}
\usepackage{amssymb}
\usepackage[dvips]{graphicx}
\usepackage{enumerate}
\usepackage{epsfig}
\usepackage{subfigure}
\usepackage{xcolor}
\usepackage[T1]{fontenc}
\usepackage{fullpage}
\usepackage{amsthm,amsfonts,amssymb,amscd,mathrsfs,xspace,framed}
\usepackage{mathrsfs,amsmath}
\usepackage{color}
\usepackage{setspace}
\usepackage{url}
\usepackage{wrapfig}
\usepackage{tikz}
\usepackage{enumitem}
\usepackage{bm}
\usepackage{comment}
\usepackage{txfonts}
\usepackage{array}
\usepackage{cancel}
\usepackage{braket}
\usepackage{mathtools}

\usepackage{amsmath}
\usepackage[acronym]{glossaries}
\usepackage[english]{babel}
\usepackage{url}
\usepackage{bm}
\usepackage{hyperref}
\definecolor{darkblue}{rgb}{0,0,0.5}
\hypersetup{
colorlinks=true,
linkcolor=black,
filecolor=blue,
citecolor=darkblue,  
urlcolor=black,
}

\usepackage[style=numeric,sorting=none]{biblatex}
\newcommand{\calE}{{\cal E}}

\begin{document}

\title{Entanglement-assisted multi-aperture pulse-compression radar for angle resolving detection}

\author{Bo-Han Wu$^{1,*}$, Saikat Guha$^{2,3}$,
Quntao Zhuang$^{2,3,\dagger}$}

\address{$^1$ Department of Physics, University of Arizona, Tucson, Arizona 85721, USA}

\address{$^2$ J. C. Wyant College of Optical Sciences, University of Arizona, Tucson, Arizona 85721, USA}
\address{$^3$ Department of Electrical and Computer Engineering, University of Arizona, Tucson, Arizona 85721, USA}

\ead{$^*$gowubohan@arizona.edu,$\dagger$zhuangquntao@arizona.edu}
\date{\today}

\begin{abstract}
Entanglement has been known to boost target detection, despite  it being destroyed by lossy-noisy propagation. Recently, [Phys. Rev. Lett. 128, 010501 (2022)] proposed a quantum pulse-compression radar to extend entanglement's benefit to target range estimation. In a radar application, many other aspects of the target are of interest, including angle, velocity and cross section.
In this study, we propose a dual-receiver radar scheme that employs a high time-bandwidth product microwave pulse entangled with a pre-shared reference signal available at the receiver, to investigate the direction of a distant object and show that the direction-resolving capability is significantly improved by entanglement, compared to its classical counterpart under the same parameter settings. We identify the applicable scenario of this quantum radar to be short-range and high-frequency, which enables entanglement's benefit in a reasonable  integration time. 
\end{abstract}


\section{Introduction}
Radio Detection and Ranging (radar) system exploits the techniques of transmitting and receiving electromagnetic (EM) field for detecting properties of distant objects. In a radar system, the ranging measurement of an object can be inferred by the time-of-flight of a pulse~[1-7]. Conventional radars use a classical coherent EM field probe. With the recent emergence of quantum sensing technology~[8-11], quantum radar has been proposed utilizing entanglement for higher sensitivity~[12-14]. The notion of quantum radar started with the detection of absence or presence of a target in the quantum illumination protocol~[15-17], where a six-decibel advantage was found in the error exponent. Recent works have also extended the applicability of quantum radar to ranging~[18,19], showing a huge entanglement advantage in range accuracy due to the nonlinear nature of the range estimation problem. However, radar detection is a complex task aiming at estimating various properties of the target, such as range, angle, velocity and cross section. The benefit of quantum radar has not been fully explored in estimating these different properties.


In this paper, we consider a bistatic dual-receiver quantum radar scheme to resolve the angular elevation of a distant target. In the large signal-to-noise ratio (SNR) limit, we identify a factor of two angle resolution advantage from entanglement. Furthermore, in the intermediate SNR region, by evaluating the quantum Ziv-Zakai bound~[20], we identify a huge angle resolution advantage from entanglement at the SNR threshold, similar to Ref.~[18]. To connect to practical scenarios, we analyze entanglement's advantage in angle detection of an unmanned aerial system (UAS) versus its range and integration time, where we identify short-range of a hundred meters to be the parameter region where entanglement's benefit over a classical radar is applicable.

\section{Radar Angle estimation}
\begin{figure}[t]
    \centering
	{\centering\includegraphics[width=0.5\linewidth]{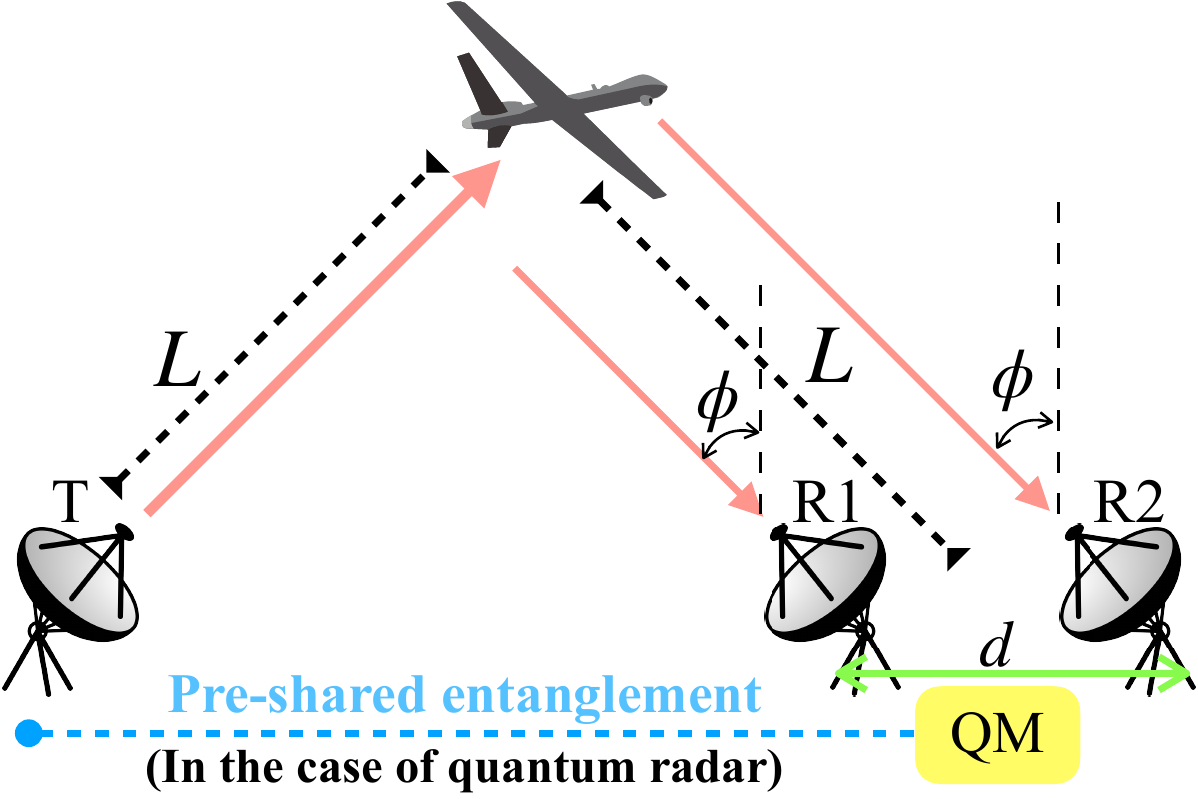}}
	\caption{Scheme of dual-receiver bistatic radar. R1: receiver 1, R2: receiver 2. T: transmitter. QM: quantum memory.}
	\label{fig:scheme}
\end{figure}

To precisely determine the angle relative to the vertical direction, $\phi$, shown in Fig.~\ref{fig:scheme},
we consider two quantum receivers, each with an individual aperture, separated by distance $d$. By assuming a distant target ($L\gg d$), the return microwave wave vectors are approximated as parallel between the two receivers. The overall return-path transmissivity (source-to-target-to-receiver) $\kappa\ll1$ is assumed to be small, and known a priori.


Our proposed quantum radar detects the angle $\phi$ by analyzing the difference of signal arrival time at the two receivers. Based on any prior knowledge of $\phi$, for example acquired using passive imaging on a classical-radar pre-estimate, we direct the dual-receivers toward $\phi_c\sim\phi$ to maximize the transmissivity of the return field. (Here `c' denotes compensation.) By doing so, the effective transmissivity, projected on the plane of the receiver, is $\kappa_{\phi,\phi_c}=\kappa\cos\left(\phi-\phi_c\right)$, which decays with the error of the prior knowledge $|\phi-\phi_c|$.

\begin{figure}[t]
    \centering
	{\centering\includegraphics[width=0.5\linewidth]{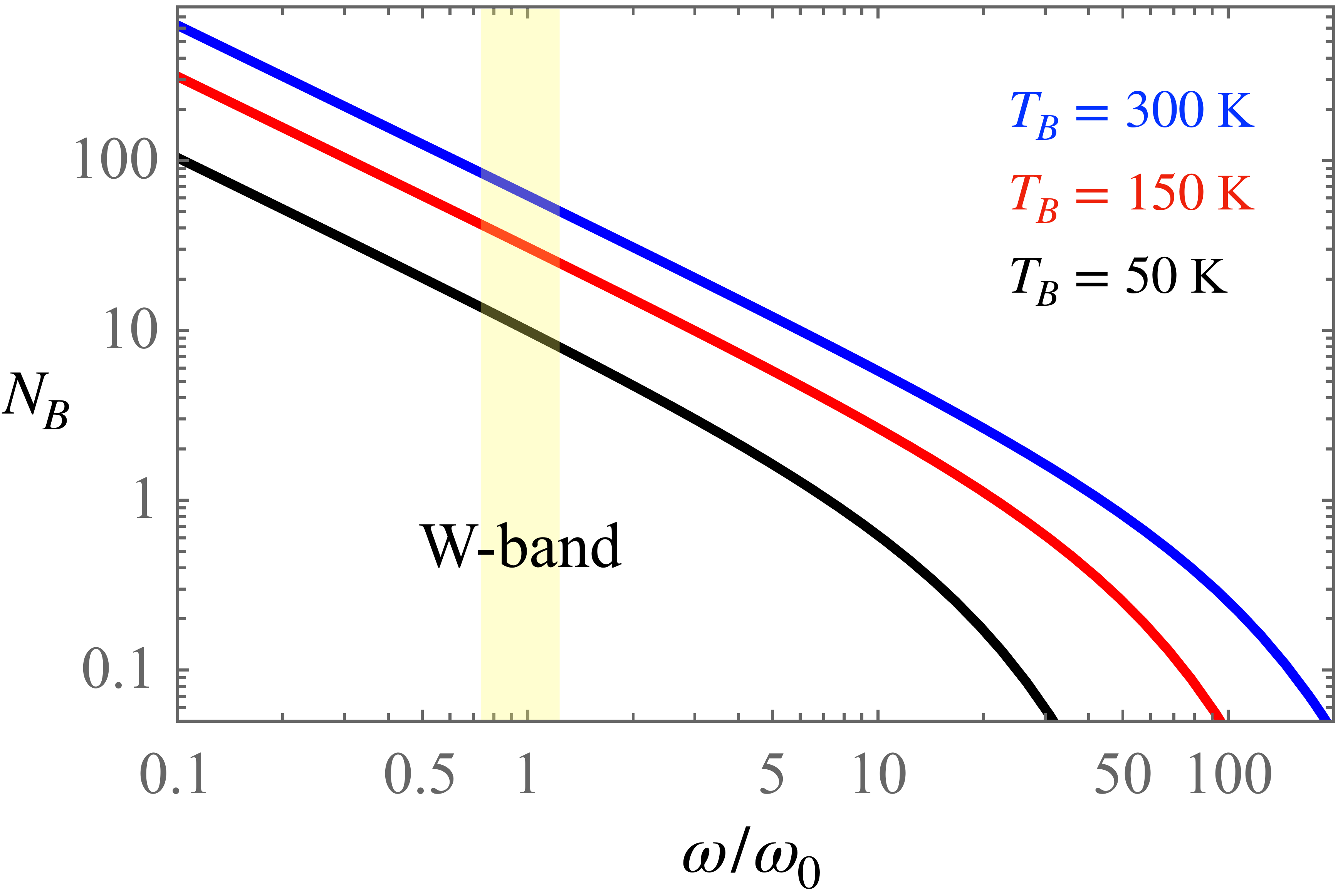}}
	\caption{Mean photon number per mode in the thermal environment, for the Planck-Law distribution.}
	\label{fig:planck}
\end{figure}

Radar, in general, is operated at microwave frequency for both quantum~[21,22] and classical~[5,6]. In our case, we choose the microwave frequency at $\omega_0/2\pi\sim 100$~GHz in W-band, since that this frequency band is robust to degraded visual environments~[23] with high precision and especially suitable for the applications of UAS localization~[24]. The thermal bath has mean photon number per mode following the Planck-law distribution, plotted in Fig.~\ref{fig:planck},
\begin{equation}
    N_B=\frac{1}{e^{\hbar\omega/k_{B}T_B}-1},
\end{equation}
where $\hbar$ is the reduced Planck constant, $k_B$ is the Boltzmann constant and $T_B$ is the temperature of the thermal bath. From this, we can see that the W-band domain is especially noisy (i.e., $N_B\gg1$). Thereby, the return-path propagation in W-band can be modeled as a very noisy and lossy channel.

Regardless of the radar scheme being classical or quantum, the input-output relation for the field operators (in units $\sqrt{\text{photons}/\text{second}}$) are the same. For transmitted field $\hat{E}\left(t\right)$, the return field at receiver 1(2) $\hat{E}_{1(2)}\left(t\right)$ can be described as
\begin{equation}
    \hat{E}_{\text{1}(2)}\left(t\right)=\sqrt{\kappa_{\phi,\phi_c}}\exp{\left(i\xi\right)}\hat{E}\left[t-\left(\tau\mp d\sin\phi/2c\right)\right]+\sqrt{1-\kappa_{\phi,\phi_c}}\hat{\mathcal{V}}_{1(2)}\left(t\right),
 \label{eq:fieldtime}
\end{equation}
where $\xi\in\left[0,2\pi\right)$ denotes the phase picked up from the reflection of the target, $\tau=2L/c$ is the time of flight of the microwave probe pulse, $c$ is the speed of light,  $\hat{\mathcal{V}}_{1}\left(t\right)$ and $\hat{\mathcal{V}}_{2}\left(t\right)$ are the environmental noise field operators that correspond to the `$-$' and `$+$' signs in `$\mp$' of Eq.~\eqref{eq:fieldtime}.
In a general phase-incoherent scenario, $\xi$ is random and unknown; to begin with, we will consider the phase-coherent case of known $\xi$, recognizing that the results obtained are lower bounds on the phase-incoherent counterparts, similar to Ref.~[18]. 
Both $\hat{E}_{1(2)}\left(t\right)$ and $\hat{\mathcal{V}}_{1(2)}\left(t\right)$ satisfy the commutation relations,
\begin{equation}
    \left[\hat{E}_{1(2)}\left(t\right),\hat{E}^{\dagger}_{1(2)}\left(t'\right)\right]=\left[\hat{\mathcal{V}}_{1(2)}\left(t\right),\hat{\mathcal{V}}^{\dagger}_{1(2)}\left(t'\right)\right]=\delta\left(t-t'\right).
\end{equation}

Subsequently, we take the Fourier transform on Eq.~\eqref{eq:fieldtime} to convert field operators into the frequency domain:
\begin{equation}
    \hat{\tilde{E}}_{\text{1}(\text{2})}\left(\omega\right)=\sqrt{\kappa_{\phi,\phi_c}}\exp{\left[i\xi^{(\mp)}_{\phi,\omega}\right]}\hat{\tilde{E}}\left(\omega\right)+\sqrt{1-\kappa_{\phi,\phi_c}}\hat{\tilde{\mathcal{V}}}_{1(2)}\left(\omega\right),
    \label{eq:inoutrela}
\end{equation}
where $\xi^{\mp}_{\phi,\omega}=\xi-\omega\left(\tau\mp d\sin\phi/2c\right)$,
\begin{equation}
\begin{aligned}
        \hat{E}\left(t\right)&=\frac{1}{2\pi}\int\hat{\tilde{E}}\left(\omega\right)e^{-i\omega t}d\omega,\\
        \hat{E}_{1(2)}\left(t\right)&=\frac{1}{2\pi}\int\hat{\tilde{E}}_{1(2)}\left(\omega\right)e^{-i\omega t}d\omega,\\
        \hat{\mathcal{V}}_{1(2)}\left(t\right)&=\frac{1}{2\pi}\int\hat{\tilde{\mathcal{V}}}_{1(2)}\left(\omega\right)e^{-i\omega t}d\omega.
\end{aligned}
\end{equation}
The noise mode $\hat{\tilde{\mathcal{V}}}_{1,2}\left(\omega\right)$  (in units of $\sqrt{\text{photons}/\text{Hz}}$) satisfy the auto-correlation relations:
\begin{equation}
    \langle\hat{\tilde{\mathcal{V}}}^\dagger_{1,2}\left(\omega\right)\hat{\tilde{\mathcal{V}}}_{1,2}\left(\omega'\right)\rangle=2\pi N_B\;\delta\left(\omega-\omega'\right).
    \label{eq:auto}
\end{equation}


For simplicity, in the following section, we model each transmitted temporal mode of the microwave probe for the classical and the quantum radars respectively, as a coherent state and a two-mode squeezed vacuum (TMSV) state. Since the aforementioned two states are both Gaussian and can be fully characterized by their quadrature mean and covariance matrices (CMs), we analyze the two radars with their corresponding mean and CMs. Note that although we have introduced the continuous-time field operators to describe the radar signals, in a finite-time analysis one can always discretize the field into the orthogonal frequency modes, each with a finite frequency bin size~[18]. We will adopt this discrete-mode approach, where finite dimensional CMs are well-defined, for evaluating various quantities. Denote the number of frequency bins as $N$. The matrix elements of the CM, $\bold{V}$, of an $N$-mode Gaussian state are given by:
\begin{equation}
    \left[\bold{V}\right]_{j,l}\equiv\langle\hat{\bold{x}}_j\hat{\bold{x}}_l\rangle-\langle\hat{\bold{x}}_j\rangle\langle\hat{\bold{x}}_l\rangle,\;\;\forall j,l=\left\{1,2,\cdots,N\right\},
    \label{eq:cm}
\end{equation}
where $\hat{\bold{x}}=\left\{\hat{q}_1,\hat{p}_1,\hat{q}_2,\hat{p}_2\cdots,\hat{q}_N,\hat{p}_N\right\}^T$ with quadrature operator $\hat{q}_s=\hat{a}_s+\hat{a}_s^{\dagger}$ and $\hat{p}_s=\left(\hat{a}_s-\hat{a}_s^{\dagger}\right)/i$, $\forall s\in\left\{1,2,\cdots,N\right\}$. Here $\hat{a}_s$'s are the annihilation operators of the modes.

\subsection{Classical radar}
Classical radar transmits a coherent state $|\sqrt{\calE}s\left(t\right)e^{-i\omega_0t}\rangle$ in a compressed chirped pulse,
\begin{equation}
    s\left(t\right)=\left(2\pi/T_d^2\right)^{-1/4}\exp{\left[i\Delta\omega^2/2T_d-t^2/4T_d^2\right]},
    \label{eq:chirppule}
\end{equation}
where $\calE$ is the total mean photon number, $T_d$ is the pulse duration (i.e., $2\pi/T_d\ll\Delta\omega\ll\omega_0$) and $\Delta \omega$ is the bandwidth. The pulse has the spectrum
\begin{equation}
\begin{aligned}
    S\left(\Omega\right)=\int_{-\infty}^{\infty}s\left(t\right)e^{-i\Omega t}dt\simeq \left(\Delta\omega^2/2\pi\right)^{-1/4}\exp{\left[-\frac{\Omega^2}{4\Delta\omega^2}\right]}
    \label{eq:spectrum}
\end{aligned}
\end{equation}
with 
\begin{equation}
    \frac{1}{2\pi}\int_{-\infty}^{\infty}\Omega^2|S\left(\Omega\right)|^2d\Omega\simeq\Delta\omega^2
    \label{eq:bandwidth_classical}.
\end{equation}

Considering the received fields at the two receivers $\hat{E}_{1}$ and $\hat{E}_{2}$, after discretizing each field into $N$ frequency modes, the global system has $2N$ modes. The overall quantum state $\hat{\rho}_\text{C}^\phi$ of the system is specified by its mean and CM, with the CM given by a direct sum of each $4$-by-$4$ CM of two received modes at the same frequency, namely $\bold{V}_\text{C}=\bigoplus_{\Omega}\left(2N_B+1\right)\bold{I}_{4}$, and quadrature mean $\langle\hat{\bold{q}}\rangle^{\phi}_\text{C}=\bigoplus_{\Omega}\langle\hat{\bold{q}}\rangle^{\phi,\Omega}_\text{C}$. Here `$\bigoplus_\Omega$' is the direct sum of over all frequencies and $\bold{I}_4$ is the $4\times4$ identity matrix. The quadrature mean can be obtained as $\langle\hat{\bold{q}}\rangle^{\phi,\Omega}_\text{C}=\sqrt{2S^{(n)}\left(\Omega\right)\kappa_{\phi,\phi_c}}\left(\cos\xi^{(-)}_{\phi},\sin\xi^{(-)}_{\phi},\cos\xi^{(+)}_{\phi},\sin\xi^{(+)}_{\phi}\right)^T$, where
\begin{equation}
    S^{(n)}\left(\Omega\right)= \sqrt{2\pi}N_S\exp{\left[-\Omega^2/2\Delta\omega^2\right]}
    \label{eq:NsW}
\end{equation}
is the mean photon number per mode of the flat-top spectral mode with width $1/T_d$ centered at $\omega_0+\Omega$. The total mean photon number is 
\begin{equation}
\begin{aligned}
    \int^{T_d}_{0}dt\langle\hat{E}^\dagger\left(t\right)\hat{E}\left(t\right)\rangle &\simeq \frac{T_d}{2\pi}\int^{\infty}_{-\infty}d\Omega\, S^{(n)}\left(\Omega\right)=N_S\Delta\omega T_d\equiv \calE.
\end{aligned}
    \label{eq:meanphoton_classical}
\end{equation}


\subsection{Quantum Radar}
In a quantum radar, the transmitted microwave pulse is entangled with an idler pulse, which is stored in a quantum memory at the location of the two receivers. As the signal pulse is returned from the target and to receiver 1 and 2, we perform a joint measurement on the quantum state of the idler and return from both receivers. 

The signal (idler) field of the microwave pulse has the field operator centered at the frequency $\omega_0$ as,
\begin{equation}
        \hat{E}_{S(I)}\left(t\right)=\frac{1}{2\pi}\int\hat{\tilde{A}}_{S(I)}\left(\Omega\right)e^{-i\left(\omega_0\pm\Omega\right)t}d\Omega,
        \label{eq:SEoperator}
\end{equation}
where the `$-$' in the `$\pm$' corresponds to the idler and we have denoted $\hat{\tilde{A}}_{S(I)}\left(\Omega\right)\equiv \hat{\tilde{E}}_{S(I)}\left(\omega_0+\Omega\right)$ for simplicity. The field operators in
Eq.~\eqref{eq:SEoperator} have spectral auto-correlations 
\begin{equation}
\begin{aligned}
    \langle\hat{\tilde{A}}_{S}^\dagger\left(\Omega\right)\hat{\tilde{A}}_{S}\left(\Omega'\right)\rangle&=\langle\hat{\tilde{A}}_{I}^\dagger\left(\Omega\right)\hat{\tilde{A}}_{I}\left(\Omega'\right)\rangle=2\pi S^{(n)}\left(\Omega\right)\delta\left(\Omega-\Omega'\right),
\end{aligned}
    \label{eq:autocorrelation}
\end{equation}
and cross-correlations,
\begin{equation}
    \langle\hat{\tilde{A}}_{S}^\dagger\left(\Omega\right)\hat{\tilde{A}}_{I}\left(\Omega'\right)\rangle=\langle\hat{\tilde{A}}_{I}^\dagger\left(\Omega\right)\hat{\tilde{A}}_{S}\left(\Omega'\right)\rangle=2\pi S^{(p)}\left(\Omega\right)\delta\left(\Omega-\Omega'\right),
    \label{eq:cross}
\end{equation}
where $S^{(p)}\left(\Omega\right)=\sqrt{S^{(n)}\left(\Omega\right)\left(S^{(n)}\left(\Omega\right)+1\right)}$. The total mean photon number of the signal (idler) mode is,
\begin{equation}
    \int^{T_d}_{0}dt\,\langle\hat{E}^\dagger_{S\left(I\right)}\left(t\right)\hat{E}_{S\left(I\right)}\left(t\right)\rangle =\frac{T_d}{2\pi}\int^{\infty}_{-\infty}d\Omega\, S^{(n)}\left(\Omega\right)=N_S\Delta\omega T_d\equiv \calE,
    \label{eq:meanphoton}
\end{equation}
and its average bandwidth is
\begin{equation}
    \int_{-\infty}^{\infty}S^{(n)}\left(\Omega\right)\Omega^2d\Omega/\int_{-\infty}^{\infty}S^{(n)}\left(\Omega\right) d\Omega=\Delta\omega^2.
    \label{eq:bandwidth}
\end{equation}
Eq.~\eqref{eq:meanphoton} and Eq.~\eqref{eq:bandwidth} coincide with the mean photon number in Eq.~\eqref{eq:meanphoton_classical} and the bandwidth in Eq.~\eqref{eq:bandwidth_classical} of the classical framework, showing that the power and the bandwidth of our quantum-radar transmitter are identical with the classical case.

After the discretization, the overall received field in the quantum radar case can be described by a collection of $N$ mode triplets, each triplet consisting of one idler mode and two received modes at receiver 1 and receiver 2. The global state $\hat{\rho}_\text{Q}^\phi$ is zero-mean Gaussian state characterized by the CM $\bold{V}^{\phi}_\text{Q}=\bigoplus_{\Omega}\bold{V}_\text{Q}^{\phi,\Omega}$, where
\begin{equation}
\begin{aligned}
    \bold{V}^{\phi,\Omega}_\text{Q}=\begin{pmatrix}
    A_\Omega\bold{I}_2&C^{\Omega}_{\phi}\bold{R}_{\phi^{(-)}_{\Omega}}&C^{\Omega}_{\phi}\bold{R}_{\phi_{\Omega}^{(+)}}\\
    C^{\Omega}_{\phi}\bold{R}_{\phi^{(-)}_{\Omega}}&B^{\Omega}_{\phi}\bold{I}_2&D^{\Omega}_{\phi}\bold{W}_{\phi_{\Omega}^{(+)}-\phi_{\Omega}^{(-)}}
    \\C^{\Omega}_{\phi}\bold{R}_{\phi^{(+)}_{\Omega}}&D^{\Omega}_{\phi}\bold{W}_{\phi_{\Omega}^{(-)}-\phi_{\Omega}^{(+)}}&B^{\Omega}_{\phi}\bold{I}_{2}
    \end{pmatrix}
\end{aligned}
\end{equation}
with $A_\Omega=2S^{(n)}\left(\Omega\right)+1$, $B^{\Omega}_{\phi}=2N_B+2\kappa_{\phi,\phi_c} S^{(n)}\left(\Omega\right)+1$, $C^{\Omega}_{\phi}=2\sqrt{\kappa_{\phi,\phi_c}}S^{(p)}\left(\Omega\right)$, $D^{\Omega}_{\phi}=A_\Omega\kappa_{\phi,\phi_c}$, $\bold{R}_\theta=\text{Re}\left[\exp{\left(i\theta\right)}\left(\bold{Z}_2-i\bold{X}_2\right)\right]$, $\bold{W}_\theta=\text{Re}\left[\exp{\left(i\theta\right)}\left(\bold{I}_2+\bold{Y}_2\right)\right]$. Here $\bold{I}_2$, $\bold{X}_2$, $\bold{Y}_2$ and $\bold{Z}_2$ are the $2\times2$ identity, Pauli X, Y, and Z matrices, $\phi_{\Omega}^{(\pm)}=\xi+\left(\omega_0+\Omega\right)\left(\tau\pm d\sin\phi/2c\right)$, $N_B=\langle\hat{\mathcal{V}}_{1(2)}^{\dagger}\hat{\mathcal{V}}_{1(2)}\rangle/\left(1-\kappa\cos\phi\right)\simeq\langle\hat{\mathcal{V}}_{1(2)}^{\dagger}\hat{\mathcal{V}}_{1(2)}\rangle$.

\subsection{Variance bound of estimator}
In this section, we evaluate lower bounds of the mean squared error (MSE) in angle estimation.
The Cram\'er-Rao bound (CRB) provides an asymptotically tight lower bound on the minimum possible MSE among all unbiased estimators. Whereas CRB is well known to be tight in the limit of infinite time-bandwidth product (i.e., $\Delta\omega T_d\rightarrow\infty$), it underestimates the achievable error as the time-bandwidth product is finite (i.e., $\Delta\omega T_d<\infty$). The Ziv-Zakai bound (ZZB) is another lower bound, obtained by analyzing the error probability in a binary hypothesis problem, and is proved to be a tighter bound than CRB in multiple cases of the non-asymptotic region~[18,20,25-30]. In the following, we compare the estimation of $\phi$ by CRBs and ZZBs in the quantum and classical cases.

\subsubsection{Cram\'er-Rao bound}
\label{sec:CRB}
In radar detection, the target angle $\phi$ is encoded into the output state $\hat{\rho}_{\phi}$. CRB indicates the minimum variance lower bound of the unbias estimator for estimating $\phi$ from $\hat{\rho}_{\phi}$, $\delta\phi_{\text{CRB}}^2\equiv1/\mathcal{F}_\phi$, where $\mathcal{F}_\phi$ is the quantum fisher information (QFI),
\begin{equation}
    \mathcal{F}_\phi\equiv\lim_{\epsilon\rightarrow0}\;8\frac{1-\sqrt{F\left(\hat{\rho}_{\phi},\hat{\rho}_{\phi+\epsilon}\right)}}{\epsilon^2},
    \label{eq:qfi}
\end{equation}
and 
\begin{equation}
    F\left(\hat{\rho}_{\phi},\hat{\rho}_{\phi+\epsilon}\right)=\left[\text{Tr}\left(\sqrt{\sqrt{\hat{\rho}_{\phi}}\hat{\rho}_{\phi+\epsilon}\sqrt{\hat{\rho}_{\phi}}}\right)\right]^2.
    \label{eq:QFid}
\end{equation}
is the Uhlmann fidelity~[31] between states $\hat{\rho}_{\phi}$ and $\hat{\rho}_{\phi+\epsilon}$.
For our evaluation, in the classical scenario, we assign $\left\{\hat{\rho}_{\phi},\hat{\rho}_{\phi+\epsilon}\right\}=\left\{\hat{\rho}^{\phi}_\text{C},\hat{\rho}^{\phi+\epsilon}_\text{C}\right\}$, whereas in quantum scenario, $\left\{\hat{\rho}^{\phi}_\text{Q},\hat{\rho}^{\phi+\epsilon}_\text{Q}\right\}$. Under the approximation $N_S,\kappa\ll1$ and $N_B\gg1$, we have the fidelities of classical and quantum cases as $F^\Omega_{\text{C},\phi,\epsilon}\simeq1-\kappa S^{(n)}\left(\Omega\right)\Theta^\Omega_{\phi,\epsilon}/N_B$ and $F^\Omega_{\text{Q},\phi,\epsilon}\simeq1-2\kappa S^{(n)}\left(\Omega\right)\Theta^\Omega_{\phi,\epsilon}/N_B$ at frequency $\omega_0+\Omega$, where 
\begin{equation}
\begin{aligned}
        \Theta^\Omega_{\phi,\epsilon}\equiv&\cos\left(\phi-\phi_c\right)+\cos\left(\phi-\phi_c+\epsilon\right)\\
        &-2\sqrt{\cos\left(\phi-\phi_c\right)\cos\left(\phi+\epsilon-\phi_c\right)}\cos\left[\frac{\left(\omega_0+\Omega\right) d}{2c}\left\{\sin\left(\phi+\epsilon\right)-\sin\phi\right\}\right].
\end{aligned}
    \label{eq:qcb}
\end{equation}

Followed by the definition of QFI in Eq.~\eqref{eq:qfi}, we derive the QFI for classical and quantum radar as,
\begin{equation}
        \mathcal{F}^{\Omega}_{\text{C},\phi}\simeq \frac{1}{2}\mathcal{F}^{\Omega}_{\text{Q},\phi}
        \simeq\frac{S^{(n)}\left(\Omega\right)\kappa_{\phi,\phi_c}}{N_B}\left\{\frac{d^2\left(\omega_0+\Omega\right)^2}{c^2}\cos^2\phi+\tan^2\left(\phi-\phi_c\right)\right\}.
        \label{eq:QFIomega}
\end{equation}
Since QFI is additive across all the frequency modes $\Omega$, we integrate Eq.~\eqref{eq:QFIomega} over the whole fluorescence spectrum. This can be justified by taking a continuous limit of a discrete set of frequency modes~[18]; in other words, we calculate $\mathcal{F}_{\text{C}(\text{Q}),\phi}= T_d\int_{-\infty}^{\infty}\mathcal{F}_{\text{C}(\text{Q}),\phi}^{\Omega}d\Omega$ to attain the gross QFIs of both scenarios $\mathcal{F}_{\text{C},\phi}\simeq2\left(\text{SNR}\right)\Upsilon_\phi$ and $\mathcal{F}_{\text{Q},\phi}\simeq4\left(\text{SNR}\right)\Upsilon_\phi$, where $\text{SNR}\equiv \kappa\mathcal{E}/N_B=\Delta\omega T_d \kappa N_S/N_B$ and \begin{equation}
        \Upsilon_\phi\equiv\frac{\cos\left(\phi-\phi_c\right)}{2}\left\{\frac{d^2\left(\omega^2_0+\Delta\omega^2\right)}{c^2}\cos^2\phi+\tan^2\left(\phi-\phi_c\right)\right\}.
    \label{eq:Upsilon}
\end{equation} 
It is interesting to note that the QFI of quantum radar is twice of that of the classical radar, similar to the ranging case~[18]. Owing to the fact that CRB is only tight in large SNR limit, we take $\phi_c=\phi$ for evaluating the CRB. Ultimately, we derive the quantum CRB (QCRB) and classical CRB (CCRB) as $\delta\phi_{\text{CCRB}}^2=1/\mathcal{F}_{\text{C},\phi}$ and $\delta\phi_{\text{QCRB}}^2=1/\mathcal{F}_{\text{Q},\phi}$, and plot them as the red and blue dotted lines in Fig.~\ref{fig:ZZB}.

\subsubsection{Ziv-Zakai bound}
\begin{figure}[t]
    \centering
	{\centering\includegraphics[width=0.6\linewidth]{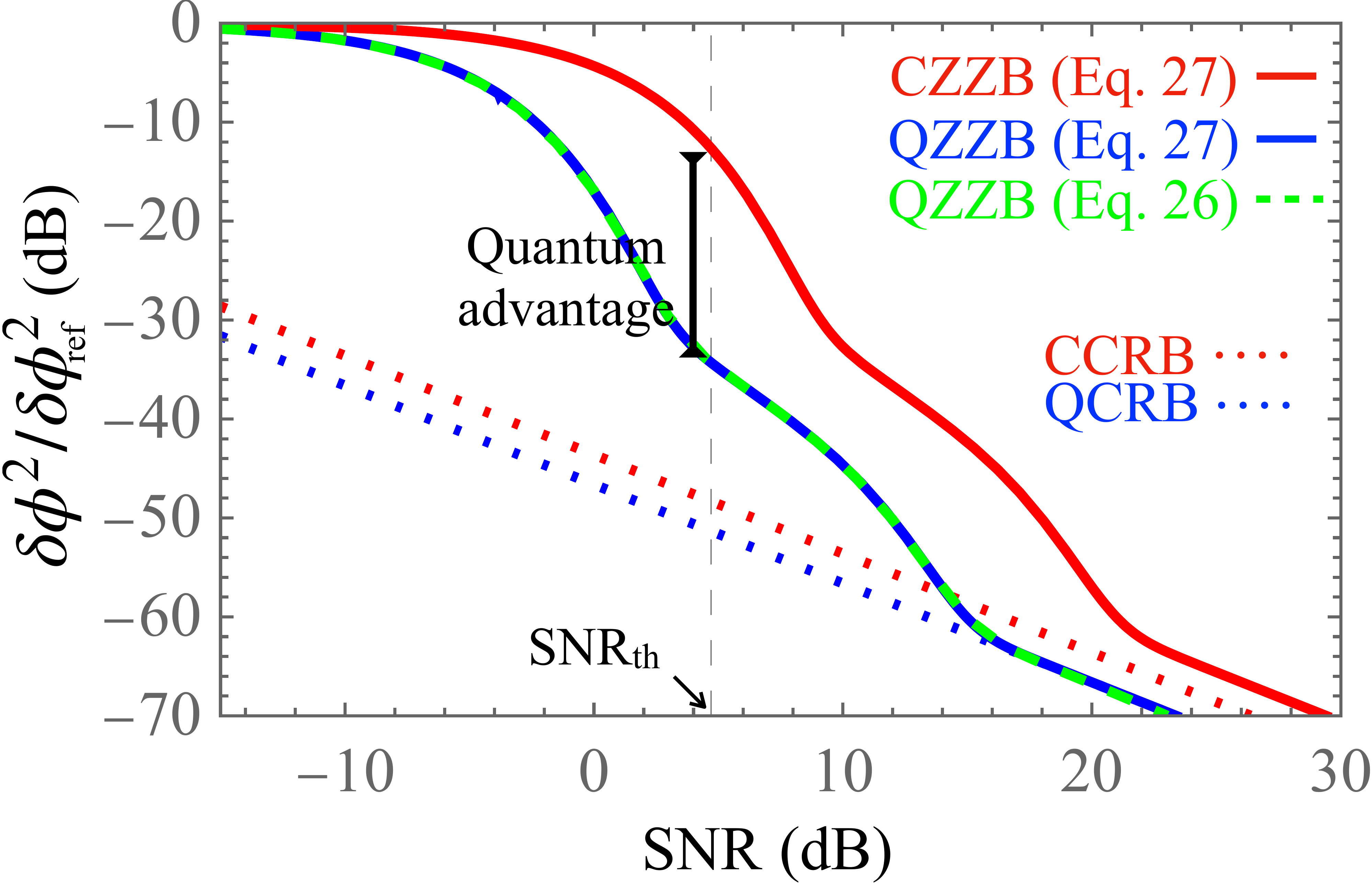}}
	\caption{The ZZBs and CRBs of quantum and classical cases normalized to $\delta\phi^2_{\text{ref}}=\Delta\phi^2/12$. Blue solid and green dashed curves denote the QZZBs derived from Eq.~\eqref{eq:zzb2} and Eq.~\eqref{eq:zzb3}, the red solid curve is derived from Eq.~\eqref{eq:zzb3}, red and blue dotted lines stands for the CCRB and QCRB.  $\omega_0/2\pi=100$~GHz, $\Delta\omega/2\pi=5$~GHz, $d=8$~m, $\phi_c=\phi=0.1$~rad and $\Delta\phi=\pi/100$. The gray vertical dashed line denotes the location of SNR threshold.}
	\label{fig:ZZB}
\end{figure}

ZZB is a Bayesian bound, calculated by averaging the MSE over a priori probability density function of the estimating parameter so as to incorporate knowledge of a priori parameter space~[26,27]. Specifically, consider a random variable $X$ with the prior distribution $P_{X}\left(\cdot\right)$, and then the ZZB of the MSE can be evaluated as
\begin{equation}
\begin{aligned}
    \delta\phi^2_{\text{ZZB}}&\equiv\int_{0}^{\infty}d\zeta\,\zeta V\left\{\int_{-\infty}^{\infty}dx\,\min\left\{P_{X}\left(x\right),P_{X}\left(x+\zeta\right)\right\}\text{Pr}\left(x;x+\zeta\right)\right\},
\end{aligned}
    \label{eq:zzb} 
\end{equation}
where $P_X\left(x\right)$ denotes the probability density of prior knowledge at angle $x$, $V$ denotes the valley-filling operation, i.e., $Vf\left(\tau\right)=\max_{\eta\ge0}f\left(\tau+\eta\right)$, and Pr$\left(x;x+\zeta\right)$ denotes the minimum error probability to distinguish two hypotheses, 
\begin{equation}
   \mathcal{H}_1:X=x,\;\;\;\;\mathcal{H}_2:X=x+\zeta.
\label{eq:hypothesis}
\end{equation}
Considering uniform prior-knowledge in the range $[\phi-\Delta \phi/2, \phi+\Delta \phi/2]$, where $\Delta\phi$ denotes the uncertainty range, we rewrite Eq.~\eqref{eq:zzb} as
\begin{equation}
    \delta\phi^2_{\text{ZZB}}=\int_{0}^{\Delta\phi}d\zeta\,\zeta \left\{\frac{1}{\Delta\phi}\int_{\phi-\Delta\phi/2}^{\phi+\Delta\phi/2-\zeta}dx\,\text{Pr}\left(x;x+\zeta\right)\right\}.
    \label{eq:zzb2}
\end{equation}
Note that as $\text{Pr}\left(x;x+\zeta\right)$ decreases with $\zeta$ increasing, and the distribution is uniform, so the principle value is always achieved at $\zeta$. Considering the limit of $\Delta\phi\ll1$, we can approximate the above results as
\begin{equation}
\begin{aligned}
    \delta\phi^2_{\text{ZZB}}&\simeq\int_{0}^{\Delta\phi}d\zeta\,\zeta\left(1-\frac{\zeta}{\Delta\phi}\right)\text{Pr}\left(\phi;\phi+\zeta\right).
\end{aligned}
    \label{eq:zzb3} 
\end{equation}
We will justify this assumption later. The error probability $\text{Pr}\left(\cdot\right)$ in Eq.~\eqref{eq:zzb2} or Eq.~\eqref{eq:zzb3} is obtained from the maximum likelyhood-ratio test in classical scenario or from the Helstrom limit in quantum. While classical ZZB (CZZB) is fairly straightforward to calculate, quantum ZZB (QZZB) is challenging due to the integration of the Helstrom limit. To enable efficient evaluation, we approximate the Helstrom limit with the quantum Chernoff bound (QCB) (i.e., $\text{Pr}\rightarrow P^{(\text{QCB})}$). As QCB is exponentially tight, we expect the results to reveal the advantage of entanglement, similar to previous works~[16,18].

Similar to Refs.~[16,32,33], we can derive the QCBs for classical and quantum as (see more details in Appendix~\ref{sec:QCB})
\begin{equation}
    \begin{aligned}
        P^{(\text{QCB})}_\text{C}\left(\phi;\phi+\zeta\right)&\le\exp{\left[-\left(\text{SNR}\right)\bar{\Theta}_{\phi,\zeta}/2\right]}/2\\
        P^{(\text{QCB})}_{\text{Q}}\left(\phi;\phi+\zeta\right)&\le \exp{\left[-2\left(\text{SNR}\right)\bar{\Theta}_{\phi,\zeta}\right]}/2,
        \label{eq:analyticalQCB}
    \end{aligned}
\end{equation}
where
\begin{equation}
\begin{aligned}
    \bar{\Theta}_{\phi,\zeta}\simeq&\cos\left(\phi-\phi_c\right)+\cos\left(\phi-\phi_c+\zeta\right)-2\sqrt{\cos\left(\phi-\phi_c\right)\cos\left(\phi-\phi_c+\zeta\right)}\;\times\\
    &\exp{\left[-\frac{d^2\Delta\omega^2}{8c^2}\left\{\sin\left(\phi+\zeta\right)-\sin\phi\right\}^2\right]}\cos\left[\frac{d\omega_0}{2c}\left\{\sin\left(\phi+\zeta\right)-\sin\phi\right\}\right],
\end{aligned}
        \label{eq:batabound}
\end{equation}
under the approximation of $N_S,\kappa\ll1$ and $N_B\gg1$. In angle detection, the error exponent of QCB achieves 6-dB advantage over the classical one, same conclusion as QI. Akin to CRB evaluation, we set $\phi_c=\phi$ and plug the upper bounds of $P^{(\text{QCB})}_\text{C}$ and $P^{(\text{QCB})}_\text{Q}$ from Eq.~\eqref{eq:analyticalQCB} into Eq.~\eqref{eq:zzb3} to evaluate the ZZBs normalized to $\delta\phi^2_{\text{ref}}=\Delta\phi^2/12$ (i.e., $\delta\phi^2_{\text{ref}}$ is the ZZB by assuming $\text{Pr}=1/2$) in Fig.~\ref{fig:ZZB}. 

Under appropriate parameter settings, Fig.~\ref{fig:ZZB} shows a huge quantum advantage ($\sim30$~dB) at the SNR threshold of the quantum radar, where the precision improves drastically with SNR increasing (will be later quantified in Section~\ref{sec:QA}). In the high SNR regime, QZZB coincides with the QCRB while CZZB has an 3~dB offset higher than CCRB. Moreover, the approximation at the $\Delta\phi\ll1$ limit used in Eq.~\eqref{eq:zzb3} can be justified by the concurrence of the blue solid curve (evaluated via Eq.~\eqref{eq:zzb3}) and green dashed curve (evaluated via Eq.~\eqref{eq:zzb2}) in Fig.~\ref{fig:ZZB} when setting a small $\Delta\phi$ (e.g., $\Delta\phi=\pi/100$). 

\subsection{Quantum advantage versus pulse duration and range}
\label{sec:QA}

To understand the practical use scenario of quantum radar, it is necessary to evaluate the trade-off between the quantum advantage with the pulse duration at a given set of physical parameter (e.g., $\kappa$ and $N_B$). To calculate the ZZB without considering the long pulse approximation, we have to calculate the QCB numerically rather than adopting the asymptotic formula in Eq.~\eqref{eq:analyticalQCB}. At the same time we will adopt the full numerical approach in Eq.~\eqref{eq:zzb2}, instead of the approximated result in Eq.~\eqref{eq:zzb3}. 



We will focus on the SNR threshold of the quantum radar to evaluate the quantum advantage.
Before proceeding to the evaluation, we make our definition of SNR threshold precise.
The intermediate SNR that results in a significant drop of QZZB is defined as the SNR threshold (SNR$_{\text{th}}$) of quantum radar, manifested in Fig.~\ref{fig:ZZB}. In the high SNR regime ($\text{SNR}\gg\text{SNR}_\text{th}$), the major contribution of the integration over $\zeta$ in Eq.~\eqref{eq:zzb3} comes from the values near the origin and, thereafter, QZZB can be asymptotically derived as
\begin{equation}
    \delta\phi^2_{\text{QZZB}}\simeq1/4\Upsilon_\phi\left(\text{SNR}\right),
    \label{eq:highSNR}
\end{equation}
where $\Upsilon_\phi$ was defined in Eq.~\eqref{eq:Upsilon}. On the other hand, the QZZB in the low SNR regime ($\text{SNR}\ll\text{SNR}_\text{th}$) follows the inequality,
\begin{equation}
    \delta\phi_{\text{ref}}^2\ge\delta\phi^2_{\text{QZZB}}\gtrsim\delta\phi_{\text{ref}}^2\exp{\left[-4\left(\text{SNR}\right)\right]},
    \label{eq:lowSNR}
\end{equation}
i.e., $0\le\bar{\Theta}_{\phi,\zeta}\le2$. SNR$_{\text{th}}$ is defined as the particular SNR that matches the asymptotic limit in Eq.~\eqref{eq:highSNR} and the asymptotic lower bound limit in Eq.~\eqref{eq:lowSNR} (i.e., $\delta\phi_{\text{ref}}^2\exp{\left[-4\left(\text{SNR}\right)\right]}$), and formulated as
\begin{equation}
    \text{SNR}_{\text{th}}\equiv g\left[1/\Upsilon_\phi\delta\phi_{\text{ref}}^2\right]/4,
\end{equation}
where $g\left[y\right]$ is the inverse function of $ye^{-y}$, $\forall y>1$.

\begin{figure}
    \centering
	{\centering\includegraphics[width=0.6\linewidth]{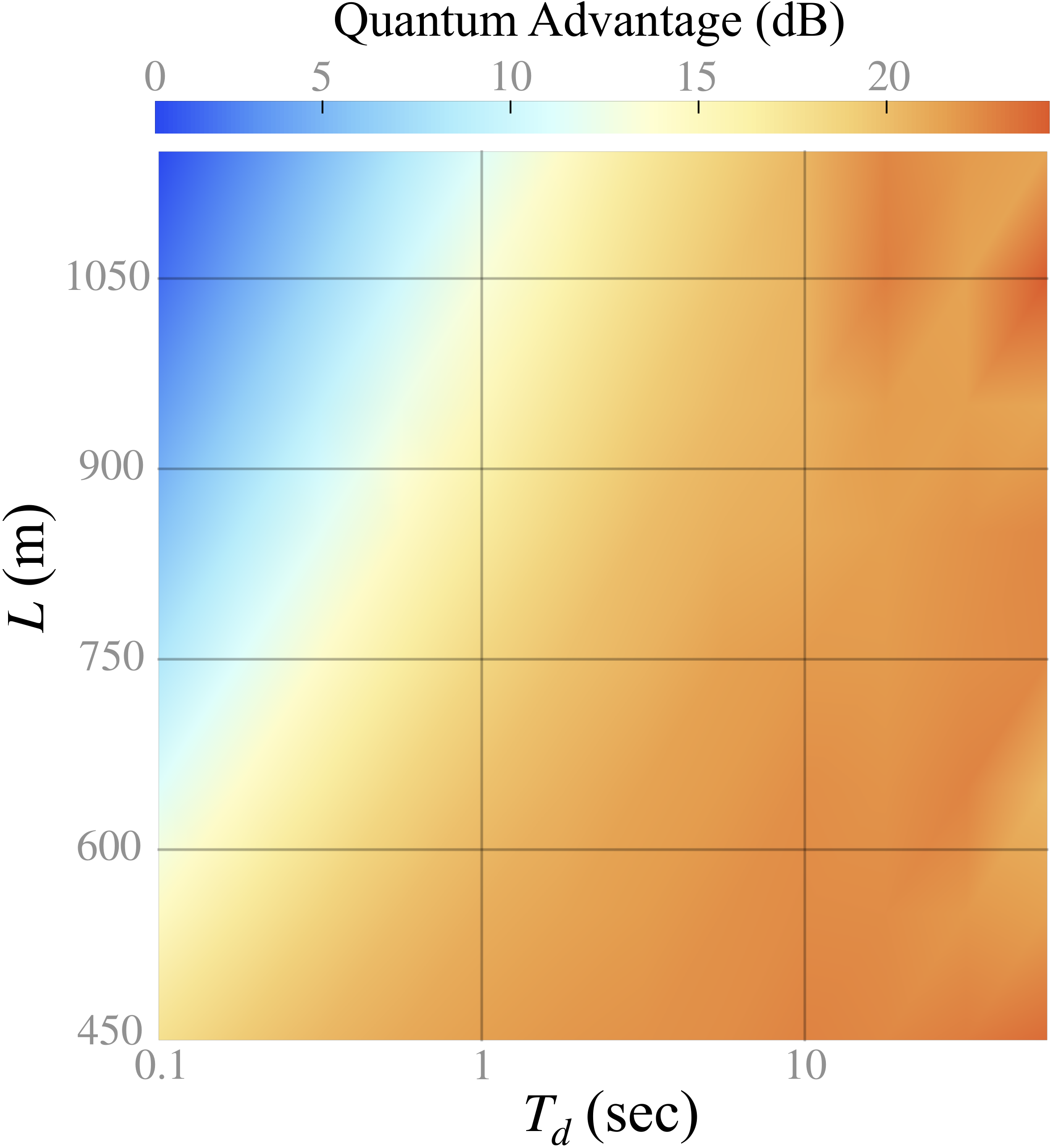}}
	\caption{The numerical calculation of quantum advantage at SNR$_{\text{th}}=4.88$~dB versus pulse duration and object's distance. $A_R=20$~m$^2$, $\sigma=0.1$~m$^2$, $N_B=32$ (i.e., $T_B=150$~K), $\omega_0/2\pi=100$~GHz, $\Delta\omega/2\pi=5$~GHz, $d=10$~m, $\phi=0.1$~rad and $\Delta\phi=\pi/100$.}
	\label{fig:2R_finite_T}
\end{figure}

To understand the quantum advantage trade-off, we specify our radar system to the application of UAS. 
In UAS detection, when the target is at distance $L$, the transmissivity of the interrogation channel 
\begin{equation}
\kappa=\frac{G_T}{4\pi L^2}\times\frac{\sigma A_R}{4\pi L^2}\ll1,
\label{kappa_radar}
\end{equation}
where $G_T=A_R/\left(2\pi c/\omega_0\right)^2$ is radar's antenna gain, $A_R$ is the antenna's area, $\sigma$ is the target's cross section area. Plugging these parameters into our simulation model and fixing $\text{SNR}=\text{SNR}_{\text{th}}$ by tuning $N_S$, we numerically calculate the quantum advantage trade off in Fig.~\ref{fig:2R_finite_T}. Under this parameter setting, the quantum advantage is appreciable ($\gtrsim15$~dB) for distance $L\sim500$~m by setting a practical pulse duration of $T_d=0.1$~second. 
 
\section{Conclusion and discussion}

In this work, we propose a two-receiver bistatic radar framework to employ a microwave probe entangled with a reference pre-shared with the receivers, in detecting the direction of the target and prove that quantum radar outperforms the classical competitor. The proposed quantum radar has the SNR threshold (i.e., SNR$_{\text{th}}$) 6~dB lower than the classical one's and, as a result, the quantum advantage is significant when we enact the radar at the SNR$_{\text{th}}$ of quantum radar at the long integration time limit. 
When the integration time is finite, the quantum advantage applies to short range precise ranging of small targets such as UAS.
Our proposed quantum radar generalizes the previous results in quantum radar ranging~[18] towards a general quantum radar detection system capable of detecting various properties of targets.

One might argue that our dual-receiver system does not make use of the spatial field distribution on the imaging plane; indeed, concatenating our analysis with spatial mode sorter and photodetection could potentially improve the estimation of $\phi$; however, we show that such a mode sorting approach only brings marginal improvement even in the best-case scenario (see more details in Appendix~\ref{app:ms}), as the usual size of aperture is small compared with the separation of the apertures. Hence, considering the level of complexity in the receiver design, it is not necessary to incorporate such processing into our design.

\ack
The project was supported by Raytheon Missiles and Defense under a corporate IRAD project led by Mark J. Meisner. QZ also acknowledges support from NSF CAREER Award CCF-2142882.

\appendix
\section{Quantum Chernoff bound}
\label{sec:QCB}
The error probability of distinguishing two hypotheses of quantum state (i.e., $\mathcal{H}_1$ and $\mathcal{H}_2$) is upper bounded by the QCB,
\begin{equation}
   P^{(\text{QCB})}\equiv\frac{1}{2}\underset{0\le s\le1}{\text{inf}}\left\{\text{Tr}\left[\hat{\rho}_{1}^s\hat{\rho}_2^{1-s}\right]\right\},
    \label{eq:QCB}
\end{equation}
where $\hat{\rho}_1$ and $\hat{\rho}_2$ are the density operators corresponding  to $\mathcal{H}_1$ and $\mathcal{H}_2$. In radar scenario, the two hypotheses refer to parameters specified in Eq.~\eqref{eq:hypothesis}, and the QCB defined in Eq.~\eqref{eq:QCB}~[32] under the approximation $\Delta\phi\ll1$ can be derived as
\begin{equation}
    P^{(\text{QCB})}\left(\phi;\phi+\zeta\right)=\prod_{\Omega}P^\Omega\left(\phi;\phi+\zeta\right),
    \label{eq:QCBformula0}
\end{equation}
with each frequency contributing 
\begin{align}
   P^{\Omega}\left(\phi;\phi+\zeta\right)=\underset{0\le s\le1}{\text{inf}}\left\{P^{\Omega,s}\left(\phi;\phi+\zeta\right)\right\},
\label{eq:QCBformula}
\end{align}
and
\begin{equation}
    P^{\Omega,s}\left(\phi;\phi+\zeta\right)=\frac{2^{N-1}\overset{N}{\underset{j=1}{\prod}} K^{\phi,\zeta}_{j,s}}{\sqrt{\det\Lambda_{\Omega,s}^{\phi,\zeta}}}\exp{\left[-\frac{1}{2}\delta \bold{v}_\Omega^{\phi,\zeta}\left(\Lambda_{\Omega,s}^{\phi,\zeta}\right)^{-1}\left(\delta \bold{v}_\Omega^{\phi,\zeta}\right)^T\right]}.
    \label{eq:QCBformula2}
\end{equation}
Here $N$ denotes the number of involved mode (i.e., $N=2$ in classical and $N=3$ in quantum), $\delta\bold{v}_\Omega^{\phi,\zeta}=\langle\hat{\bold{q}}\rangle_{\Omega}^{\phi+\zeta}-\langle\hat{\bold{q}}\rangle_{\Omega}^{\phi}$ denotes the quadrature mean difference, $G_s^{(\pm)}\left[y\right]=\sqrt{2}/\left[\left(y+1\right)^s\pm\left(y-1\right)^s\right]$, i.e., $\forall y>1$, $K^{\phi,\zeta}_{j,s}=G_s^{(-)}\left[\lambda^{\phi}_{j}\right]\times G_{1-s}^{(-)}\left[\lambda^{\phi+\zeta}_{j}\right]$, $\Lambda_{\Omega,s}^{\phi,\zeta}=\bold{C}^{\phi}_{\Omega,s}+\bold{C}_{\Omega,1-s}^{\phi+\zeta}$,
\begin{equation}
    \bold{C}^{\phi'}_{\Omega,s}\equiv\bold{S}_{\Omega}^{\phi'}\left\{\bigoplus_{j=1}^N\frac{G_s^{(+)}\left[\lambda^{\phi'}_{j}\right]}{G_s^{(-)}\left[\lambda^{\phi'}_{j}\right]}\otimes \bold{I}_2\right\}\left(\bold{S}_{\Omega}^{\phi'}\right)^T,
    \label{eq:V}
\end{equation}
$\lambda^{\phi'}_j$ is the $j$-th symplectic eigenvalue associated with the symplectic matrix $\bold{S}^{x'}_{\Omega}$ in parameter $\phi'=\left\{\phi,\phi+\zeta\right\}$. 

\begin{figure}
    \centering
	{\centering\includegraphics[width=0.5\linewidth]{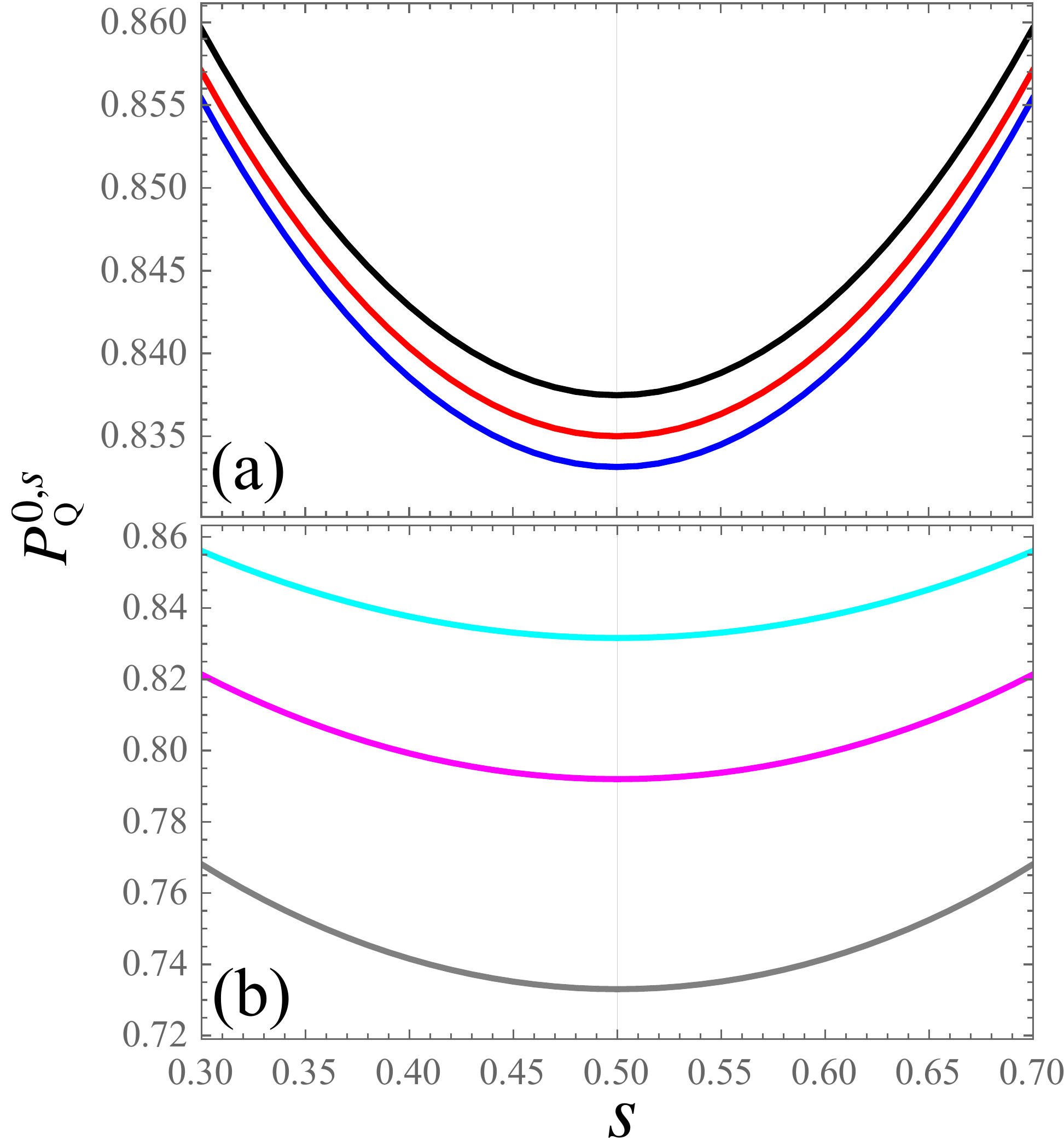}}
	\caption{$P^{0,s}_{\text{Q}}$ versus $s$ with $\omega_0/2\pi=100$~GHz, $\Delta\omega/2\pi=1$~MHz, $d=20$~m, $N_B=32$, $\phi=0$~rad and $\zeta=\pi/3$. (a) Blue, red and black curves stand for different choices of $\kappa$ and $T_d$ as $\left\{\kappa,T_d\right\}=\left\{5\times10^{-1},8~\text{s}\right\}$, $\left\{5\times10^{-3},0.8~\text{ms}\right\}$ and $\left\{5\times10^{-4},8~\text{ms}\right\}$ with $N_S=0.1$. (b) Cyan, magenta and gray curves stand for different choices of $N_S$ and $T_d$ as $\left\{N_S,T_d\right\}=\left\{2,0.32~\text{\textmu s}\right\}$, $\left\{0.63,10~\text{\textmu s}\right\}$ and $\left\{0.2,3.2~\text{\textmu s}\right\}$ with $\kappa=0.5$.} 
	\label{fig:BB}
\end{figure}

Subsequently, we claim that the infimum in Eq.~\eqref{eq:QCBformula} occurs at $s=1/2$ and support it with numerical justifications and perturbation theory. Fig.~\ref{fig:BB} shows the plot of Eq.~\eqref{eq:QCBformula2} in quantum case with $\Omega=0$ as a function of $s$ (i.e., $P^{0,s}_\text{Q}$). In Fig.~\ref{fig:BB}(a), we fix $\kappa$ while changing $N_S$; in Fig.~\ref{fig:BB}(b), $N_S$ is fixed while $\kappa$ changes. Obviously, in these parameter settings, all minimal values, consistently, occur at the choices of $s=1/2$. To be more strict on justifying $s=1/2$, we employ the perturbation theory, introduced by Ref.~[34], on our angle-resolving radar model with the approximation $N_S,\kappa\ll1$ and $N_B\gg1$, and easily conclude that infinimum does occur at $s=1/2$ by the \emph{Theorem 4} of Ref.~[34].

Therefore, we can calculate Eq.~\eqref{eq:QCBformula} in both cases as
$P^\Omega_\text{C}\left(\phi;\phi+\zeta\right)\le\exp{\left[-\kappa S^{(n)}\left(\Omega\right)\Theta^\Omega_{\phi,\zeta}/2N_B\right]}/2$ and $P^\Omega_\text{Q}\left(\phi;\phi+\zeta\right)\le\exp{\left[-2\kappa S^{(n)}\left(\Omega\right)\Theta^\Omega_{\phi,\zeta}/N_B\right]}/2$ under the approximation of $N_S,\kappa\ll1$ and $N_B\gg1$, where $\Theta^\Omega_{\phi,\zeta}$ was defined in Eq.~\eqref{eq:qcb}, yielding the QCBs in Eq.~\eqref{eq:analyticalQCB}.

\section{Mode sorter}
\label{app:ms}
\begin{figure}
    \centering
	{\centering\includegraphics[width=0.5\linewidth]{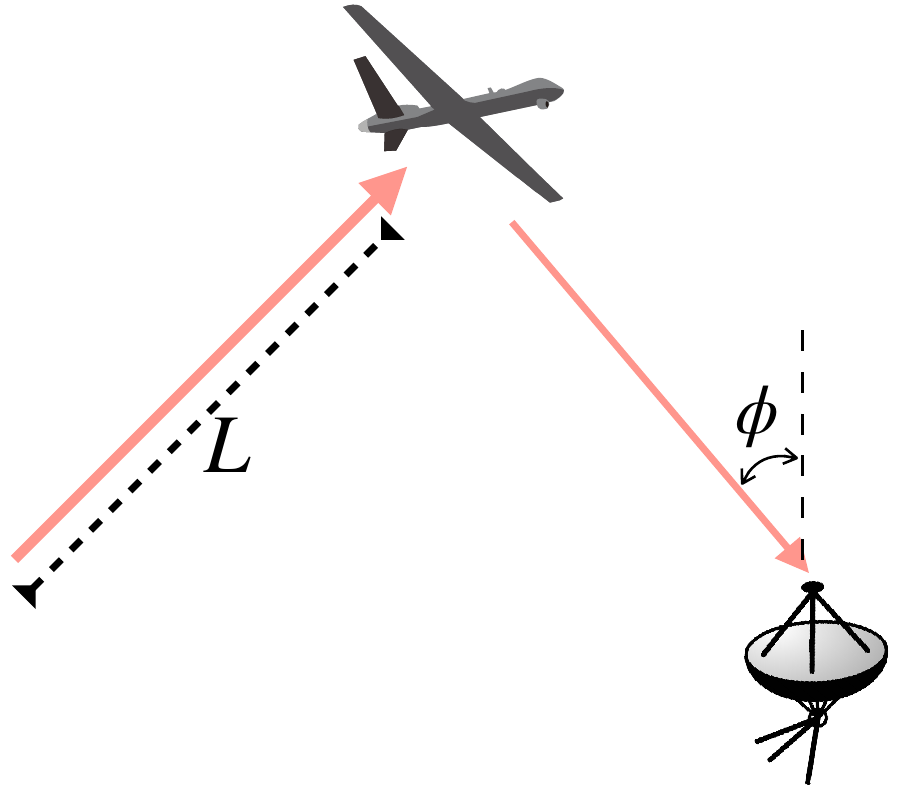}}
	\caption{The scheme of single-receiver radar.}
	\label{fig:scheme_1R}
\end{figure}
In this section, we apply mode sorter on the proposed radar and study the potential improvement. Instead of dual-receiver, let us underpin the scheme of single-receiver, shown in Fig.~\ref{fig:scheme_1R}, to simplify the calculation, and we anticipate that mode sorter brings the same or, at least, similar improvement in single- and dual-receiver schemes.

\begin{figure*}
    \centering
	{\centering\includegraphics[width=1\linewidth]{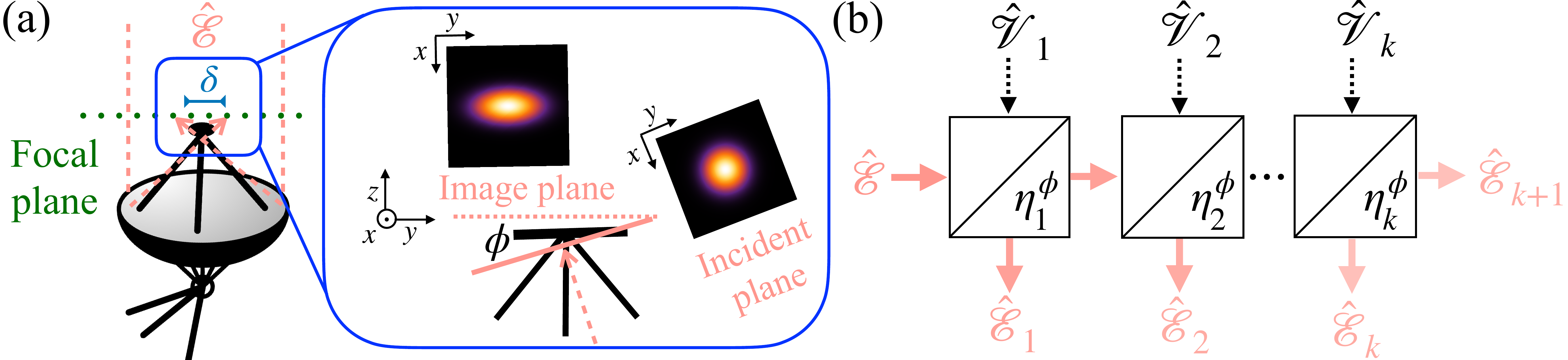}}
	\caption{Scheme of single-receiver. (a) Trajectory of receiving signal. (b) The schematics of $k$-series mode sorter.}
	\label{fig:sorter}
\end{figure*}

In a single receiver scenario, we set the target at the angle $\phi$, same as in dual-receiver radar, but set the compensation angle to be zero $\phi_c=0$ (i.e., the face of radar is vertically directed, shown in Fig.~\ref{fig:scheme_1R}) and consider a soft-aperture located at the focal plane of the paraboloid antenna in Fig.~\ref{fig:sorter}(a). The received pulse is collected by the paraboloid antenna as an elliptical Gaussian beam $w\left(t;x,y\right)=s\left( t\right)\varepsilon\left(x,y\right)$, where
\begin{equation}
    \varepsilon\left(x,y\right)=\sqrt{\frac{\cos\phi}{2\pi r^2}}\exp{\left[-\frac{x^2}{4r^2}-\frac{ y^2}{4r^2/\cos^2\phi}\right]},
    \label{eq:field1R}
\end{equation}
whose overall phase is set zero, $r$ is the half length of the minor axis of ellipse (i.e., cross section of paraboloid), and the field has the spectrum $W\left(\Omega;x,y\right)=S\left(\Omega\right)\varepsilon\left(x,y\right)$, where $s\left(t\right)$ and $S\left(\Omega\right)$ were denoted in Eq.~\eqref{eq:chirppule} and Eq.~\eqref{eq:spectrum}. The asymmetricity of the two axes comes from the angle deviation $\phi$ between the incident plane and the imaging plane, yielding the y-axis, shown in Fig.~\ref{fig:sorter}(a), being elongated whereas x-axis being the same.

Hermite Gaussian (HG) mode is treated as a discriminator to sort the spatial mode of the field on the imaging plane~[35,36], shown in Fig.~\ref{fig:sorter}(a). The proposed HG mode has the eigenfunction 
\begin{equation}
    \begin{aligned}
        \psi_{n,m}\left(x,y\right)=\frac{\beta H_m\left(\beta x\right) H_n\left(\beta y\right)}{\sqrt{\pi n!m!2^{n+m}}}\exp{\left[-\nu\beta^2\left(x^2+y^2\right)\right]},
    \end{aligned}
    \label{eq:HG}
\end{equation}
indexed by non-negative integers $n,m\in\mathbb{N}_0$, where $\beta=\sqrt{2}\left(1+4D_f\right)^{1/4}/\delta$, $D_f=\left(2\pi A_\delta c L/\omega_0\right)^2$ is the Fresnel number, $A_\delta=\pi\delta^2/4$ is the aperture area (i.e., $\delta$ is its diameter),  $\nu=\left(1+i\omega_0/c\beta^2L\right)/2\simeq1/2$, and $H_{n}\left(\cdot\right)$ denotes the $n$-th order Hermite polynomial.

The received signal can be regarded as a far field if $D_f\ll1$ and is decomposed by $k$ HG modes and an additional mode that covers the residual higher order ones. The field distribution projected on the imaging plane depends on the parameter $\phi$, as demonstrated in the inset of Fig.~\ref{fig:sorter}(a).

\begin{figure}
    \centering
	{\centering\includegraphics[width=0.5\linewidth]{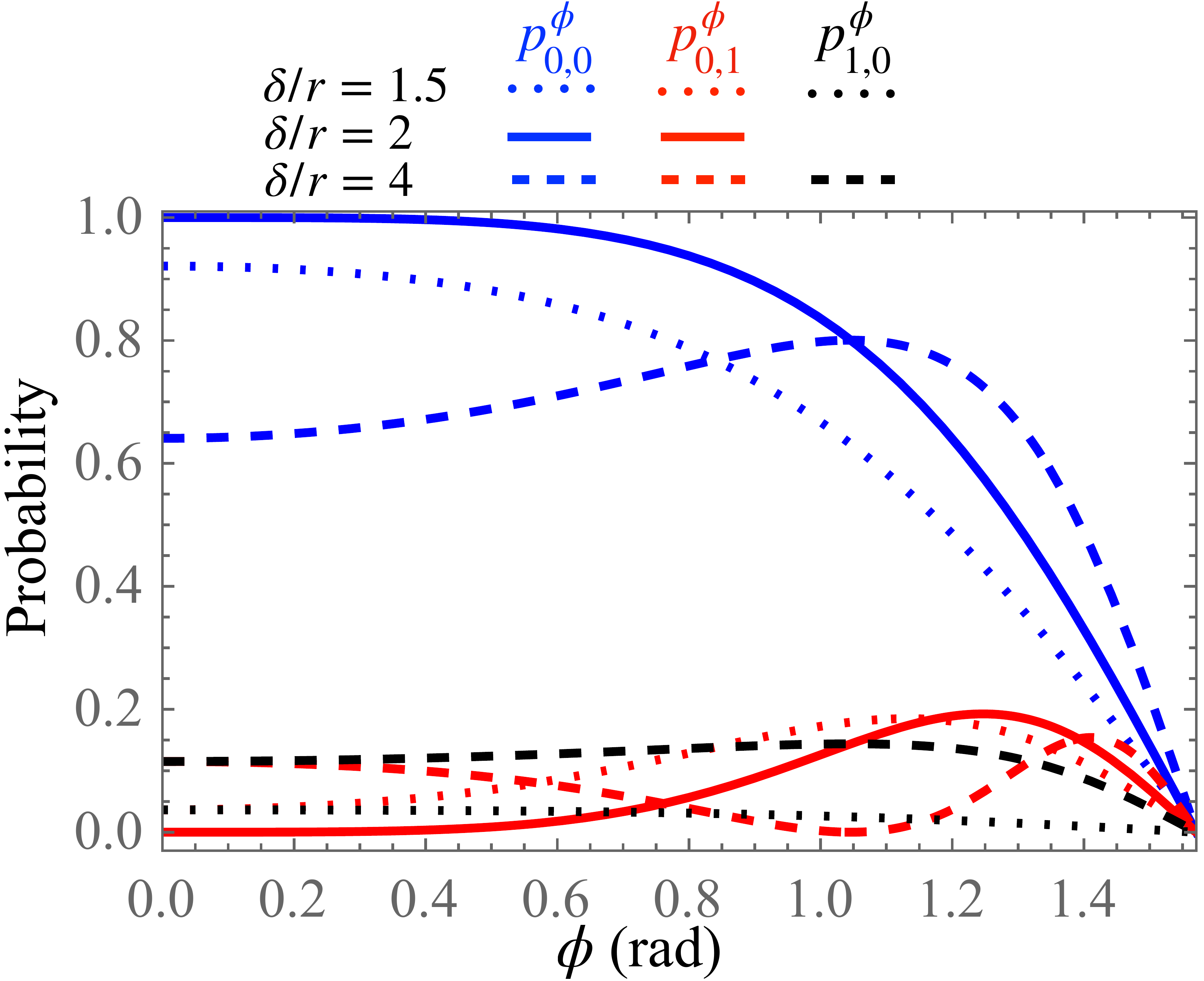}}
	\caption{Occupation probability of $\psi_{0,0}$, $\psi_{0,1}$ and $\psi_{1,0}$ in different combinations of $\delta/r$. $L=10$~km and $\omega_0/2\pi=100$~GHz.}
	\label{fig:prob}
\end{figure}

To begin our analyses, we evaluate the overlap of a Gaussian function with each basis $\psi_{n,m}\left(x,y\right)$ at the image plane, leading to the associate occupation probability
\begin{equation}
\begin{aligned}
    p_{n,m}^\phi\equiv&\left\{\frac{\left(2n-1\right)!!\left(2m-1\right)!!}{2^{n+m}n!m!}\right\}\times\frac{4\chi\cos\phi}{\left(1+\chi\right)\left(\cos^2\phi+\chi\right)}\left(\frac{1-\chi}{1+\chi}\right)^{2n}\left(\frac{\cos^2\phi-\chi}{\cos^2\phi+\chi}\right)^{2m},
\end{aligned}
\end{equation}
where $\chi=2\beta^2r^2$ and this decomposition applies to our Gaussian beam imaging system~[36]. For smooth communication, we relabel the indices of probability at each mode as $\left\{\mathcal{P}_1^\phi,\mathcal{P}_2^\phi,\mathcal{P}_3^\phi,\mathcal{P}^\phi_4,\mathcal{P}^\phi_5,\mathcal{P}^\phi_6,\cdots,\mathcal{P}^\phi_{k},1-\sum^k_{j=1}\mathcal{P}^\phi_{j}\right\}=\left\{p^\phi_{0,0},p^\phi_{1,0},p^\phi_{0,1},p^\phi_{2,0},p^\phi_{1,1},p^\phi_{0,2},\cdots\right\}$ and plot the lowest three order modes in Fig.~\ref{fig:prob}. This mathematical decomposition process can be visualized by passing the incident Gaussian beam through a $k$-series beamsplitters, shown in Fig.~\ref{fig:sorter}(b). The $j$-th mode (i.e., $1\le j\le k$) beamsplitter has the matrix form
\begin{equation}
    \bold{B}^\phi_{j}=
    \begin{pmatrix}
    \sqrt{\eta^\phi_j}&\sqrt{1-\eta^\phi_{j}}\\
    -\sqrt{1-\eta^\phi_{j}}&\sqrt{\eta^\phi_{j}}\end{pmatrix}\otimes\bold{I}_2,
\end{equation}
where 
\begin{equation}
\eta^\phi_{j}=\begin{cases}
    \mathcal{P}^\phi_j\;\;,&\;\;\;\;\;\;j=1\\
    \mathcal{P}^\phi_j/\left(1-\sum_{l=1}^{j-1}\mathcal{P}^\phi_l\right),&2\le j\le k
\end{cases}.
\end{equation}
These $k$ beamsplitters entangle the $k$ vacuum modes $\hat{\mathcal{V}}$ at the input and result in $k+1$ output modes. 

In the following, we apply the mode sorter on the analysis of classical radar and quantum radar. For simplicity, we consider the mode sorter approach that involves only one beamsplitter, $k=1$ and set $\delta/r=2$ such that the fundamental mode ($\psi_{0,0}$) dominates all other HG modes when the concerning angle is small $\phi\ll1$, shown in Fig.~\ref{fig:prob}.

\subsection{Classical radar}
\begin{figure}
    \centering
	{\centering\includegraphics[width=0.6\linewidth]{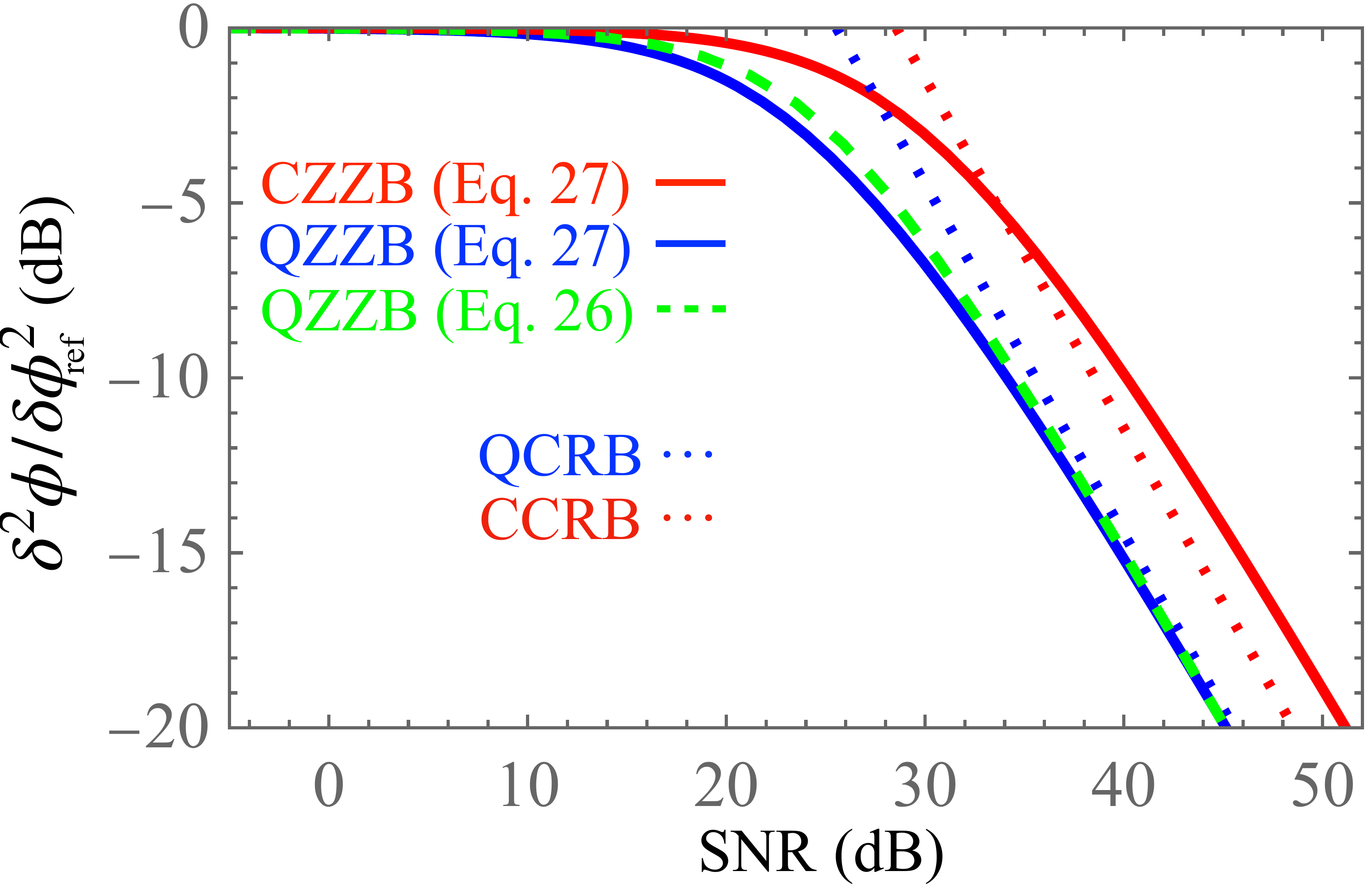}}
	\caption{The ZZBs and CRBs of quantum and classical cases in single-receiver scenario. $\Delta\phi=\pi/100$ and $\phi=\pi/2-\Delta\phi/2$.}
	\label{fig:1R}
\end{figure}

In classical single-receiver scheme, the global state, at the angle $\phi$, has the CM $\bold{V}_\text{C}=\left(2N_B+1\right)\bigoplus_{\Omega}\bold{I}_{4}$ and quadrature mean $\langle\hat{\bold{q}}\rangle_\text{C}^{\phi}=\bigoplus_{\Omega}\bold{B}^\phi_{1}\langle\hat{\bold{q}}\rangle_\text{C}^{\phi,\Omega}$, where $\langle\hat{\bold{q}}\rangle^{\phi,\Omega}_{\text{C}}=\sqrt{2S^{(n)}\left(\Omega\right)\kappa_{\phi,0}}\left(\cos\Xi,\sin\Xi,0,0\right)^T$ is the quadrature at frequency $\omega_0+\Omega$ in the basis of $\left(\hat{q}_S,\hat{p}_S,\hat{q}^{\bot}_{S},\hat{p}^{\bot}_{S}\right)$, where $\Xi\in(0,2\pi]$ is the overall phase of the return signal, $\left\{\hat{q}_S,\hat{p}_S\right\}$ and $\left\{\hat{q}^{\bot}_S,\hat{p}^{\bot}_S\right\}$ are the quadrature pairs of signal, projected on $\psi_{0,0}$ (i.e., occupation probability $\mathcal{P}^\phi_1$), and the residual HG modes $\psi^{\bot}$ (i.e., occupation probability $1-\mathcal{P}^\phi_1$). Akin to calculating CRB and ZZB of dual-receiver radar in the main text, under the approximation $N_S,\kappa\ll1,N_B\gg1$, we asymptotically and analytically derive the CCRB (red dotted line in Fig.~\ref{fig:1R}),
\begin{equation}
    \delta\phi^2_{\text{CCRB}}\simeq\left\{\frac{\text{SNR}}{2\cos\phi}\left\{\sin^2\phi+\frac{\cos^2\phi}{\mathcal{P}_1^{\phi}\left(1-\mathcal{P}_1^{\phi}\right)}\left(\frac{d\mathcal{P}_1^{\phi}}{d\phi}\right)^2\right\}\right\}^{-1}.
    \label{CCRB_classical}
\end{equation}
Similarly, the QCB for distinguishing the hypotheses in Eq.~\eqref{eq:hypothesis} can be obtained as $P^{(\text{QCB})}_{\text{C}}\left(x;x+\zeta\right)\le\exp{\left[-\left(\text{SNR}\right)\Gamma_{x,\zeta}/4\right]}/2$, where
\begin{equation}
    \Gamma_{x,\zeta}\simeq\cos x+\cos\left(x+\zeta\right)-2\sqrt{\cos\left(x\right)\cos\left(x+\zeta\right)}\left(\sqrt{\mathcal{P}^{x}_1\mathcal{P}^{x+\zeta}_1}+\sqrt{\left(1-\mathcal{P}^{x}_1\right)\left(1-\mathcal{P}^{x+\zeta}_1\right)}\right).
    \label{eq:1R_BataC}
\end{equation}
rior knowledge in $\left[\phi-\Delta\phi/2,\phi+\Delta\phi/2\right]$, CZZB is numerically derived by plugging the upper bound of $P^{(\text{QCB})}_{\text{C}}$ into Eq.~\eqref{eq:zzb3} with the approximation $\Delta\phi\ll1$ and is plotted as the red solid curve in Fig.~\ref{fig:1R}. 

In Eq.~\eqref{CCRB_classical}, note that the size of the receiver antenna comes in via the transmissivity $\kappa$ in the SNR (see Eq.~\eqref{kappa_radar}). The area of aperture $A_\delta$ at the focal plane needs to match the antenna's cross section area $A_R$ in a way to receive all the beams that is reflected from the surface of paraboloid antenna, so that no additional loss occurs to degrade the SNR.

\subsection{Quantum radar}
In a quantum single-receiver scheme, the collected microwave field is from the signal mode of a TMSV state. After HG mode decomposition, the CM of global state at angle $x'=\left\{x,x+\zeta\right\}$ is
\begin{equation}
\begin{aligned}
        \bold{V}_\text{Q}=\left(\bold{I}_2\oplus\bold{B}^{x'}_1\right)\begin{pmatrix}
        A_\Omega\bold{I}_2&\mathcal{C}^\Omega_{x'}\bold{R}_{x'}&\bold{0}_2\\
        \mathcal{C}^\Omega_{x'}\bold{R}_{x'}&\mathcal{B}^\Omega_{x'}\bold{I}_2&\bold{0}_2\\
        \bold{0}_2&\bold{0}_2&\mathcal{B}^\Omega_{x'}\bold{I}_2
    \end{pmatrix}\left(\bold{I}_2\oplus\bold{B}^{x'}_1\right)^T,
    \end{aligned}
\end{equation}
in quadrature basis $\left(\hat{q}_{I},\hat{p}_{I},\hat{q}_S,\hat{p}_S,\hat{q}^{\bot}_S,\hat{p}^{\bot}_S\right)$, where $\left\{\hat{q}_{I},\hat{p}_{I}\right\}$ denotes the quadrature pair of idler mode, $\mathcal{B}^\Omega_{x'}=2N_B+2\kappa_{x',0} S^{(n)}\left(\Omega\right)+1$, $\mathcal{C}^{\Omega}_{x'}=2\sqrt{\kappa_{x',0}}S^{(p)}\left(\Omega\right)$. The QCRB (blue dotted line in Fig.~\ref{fig:1R}) is asymptotically and analytically derived as,
\begin{equation}
    \delta\phi^2_{\text{QCRB}}\simeq\left\{\frac{\text{SNR}}{\cos\phi}\left\{\sin^2\phi+\frac{\cos^2\phi}{\mathcal{P}_1^{\phi}\left(1-\mathcal{P}_1^{\phi}\right)}\left(\frac{d\mathcal{P}_1^{\phi}}{d\phi}\right)^2\right\}\right\}^{-1}
    \label{eq:1R_BataQ}
\end{equation} and QCB, for distinguishing $\mathcal{H}_1$ and $\mathcal{H}_2$ in Eq.~\eqref{eq:hypothesis},  $P^{(\text{QCB})}_{\text{Q}}\left(x;x+\zeta\right)\le\exp{\left[-\left(\text{SNR}\right)\Gamma_{x,\zeta}\right]}/2$ under the approximation $N_S,\kappa\ll1$ and $N_B\gg1$. With uniform prior knowledge in $\left[\phi-\Delta\phi/2,\phi+\Delta\phi/2\right]$, QZZBs can be numerically calculated with or without the approximation $\Delta\phi\ll1$ by Eq.~\eqref{eq:zzb3}, which are plotted as the blue solid and green dashed curves in Fig.~\ref{fig:1R}. Same conclusion as dual-receiver, the concurrence of two curves justifies the formula of Eq.~\eqref{eq:zzb3} and they both coincide with the QCRB in high SNR regime. In Fig.~\ref{fig:1R}, we plot ZZBs in both classical and quantum scenarios by fixing the uncertainty tolerance $\Delta\phi=\pi/100$ and setting $\phi=\pi/2-\Delta\phi/2$. The choice of $\phi=\pi/2-\Delta\phi/2$ has the maximal distinguishability between $\mathcal{H}_1$ and $\mathcal{H}_2$, because the occupation probability in HG fundamental mode varies dramatically as target angle approaches $\pi/2$. However, in Fig.~\ref{fig:1R}, despite the optimal choice of $\phi$, the noticeable reduction of $\delta\phi^2/\delta\phi^2_{\text{ref}}$ can only be observed when SNR goes to very high (i.e., $>30$~dB); conversely, the dual-receiver radar mode sorter has significant reduction outcome even at low SNR regime (i.e., $\sim1$~dB).

\section*{References}
\begin{thebibliography}{100}

\bibitem{Trees01}
H. L. Van Trees, \emph{Detection, Estimation, and Modulation Theory, Part I: Detection, Estimation, and Linear Modulation Theory} (Wiley, New York, 2001).

\bibitem{Trees2001}
H. L. Van Trees, \emph{Detection, Estimation, and Modulation Theory, Part III: Radar-Sonar Signal Processing and Gaussian Signals in Noise} (Wiley, New York, 2001).

\bibitem{Mallinckrodt54}
A. Mallinckrodt and T. Sollenberger, \emph{Optimum pulse-time determination}, IRE Trans. Inf. Theory {\bf3}, 151 (1954).

\bibitem{Skolnik60}
M. I. Skolnik, \emph{Theoretical accuracy of radar measurements}, IRE transactions on aeronautical and navigational electronics ANE-7, 123 (1960).

\bibitem{Skolnik02}
M. I. Skolnik, \emph{Introduction to Radar Systems, Third Edition} (McGraw-Hill, New York, 2002).

\bibitem{Handbook08}
\emph{Radar Handbook}, edited by M. I. Skolnik (McGraw Hill, New York, 2008).

\bibitem{Marcum60}
J. Marcum, \emph{A statistical theory of target detection by pulsed radar}, IRE Trans. Inf. Theory {\bf6}, 59 (1960).

\bibitem{Giovannetti01}
V. Giovannetti, S. Lloyd, and L. Maccone, \emph{Quantum enhanced positioning and clock synchronization}, Nature
(London) {\bf412}, 417 (2001).

\bibitem{Giovannetti04}
V. Giovannetti, S. Lloyd, and L. Maccone, \emph{Quantum enhanced measurements: Beating the standard quantum limit}, Science {\bf306}, 1330 (2004).

\bibitem{Shapiro07}
J. H. Shapiro, \emph{Quantum pulse compression laser radar}, Proc. SPIE Int. Soc. Opt. Eng. {\bf6603}, 660306 (2007).

\bibitem{Maccone20}
L. Maccone and C. Ren, \emph{Quantum Radar}, Phys. Rev. Lett. {\bf124}, 200503 (2020).

\bibitem{Lanzagorta12}
M. Lanzagorta, \emph{Quantum Radar Synthesis Lectures on Quantum Computing} (Morgan, Claypool, San Rafael, 2012).

\bibitem{Torrom20}
R. G. Torrom\'e, N. B. Bekhti-Winkel, and P. Knott,
\emph{Introduction to quantum radar}, arXiv:2006.14238 (2020).

\bibitem{Shapiro20}
J. H. Shapiro, \emph{The quantum illumination story}, IEEE Trans. Aerosp. Electron. Syst. {\bf35}, 8 (2020).


\bibitem{Tan08} 
S.-H. Tan \emph{et al}, \emph{Quantum Illumination with Gaussian States}, Phys. Rev. Lett. {\bf101}, 253601 (2008).

\bibitem{Zhuang17} 
Q. Zhuang, Z. Zhang, and J. H. Shapiro, \emph{Optimum Mixed State Discrimination for Noisy Entanglement-Enhanced Sensing}, Phys. Rev. Lett. {\bf118}, 040801 (2017).


\bibitem{Lloyd08}
S. Lloyd, \emph{Enhanced sensitivity of photodetection via quantum illumination}, Science {\bf321}, 1463 (2008).

\bibitem{Zhuang22}
Q. Zhuang and J. H. Shapiro, \emph{Ultimate Accuracy Limit of Quantum Pulse-Compression Ranging}, Phys. Rev. Lett. {\bf128}, 010501 (2022).

\bibitem{Zhuang21}
Q. Zhuang, \emph{Quantum Ranging with Gaussian Entanglement}, Phys. Rev. Lett. {\bf126}, 240501 (2021).

\bibitem{Tsang12}
M. Tsang, \emph{Ziv-Zakai Error Bounds for Quantum Parameter Estimation}, Phys. Rev. Lett. {\bf108}, 230401 (2012).

\bibitem{Ebrahimi22}
M. S. Ebrahimi, S. Zippilli, D. Vitali, \emph{Feedback-enabled Microwave Quantum Illumination}, Quantum Sci. Technol. {\bf 7} 035003 (2022)


\bibitem{Barzanjeh15} 
S. Barzanjeh, S. Guha, C. Weedbrook, D. Vitali, J. H.
Shapiro, and S. Pirandola, \emph{Microwave Quantum Illumination}, Phys. Rev. Lett. {\bf114}, 080503 (2015).

\bibitem{Sanders18}
J. N. Sanders-Reed and S. J. Fenley, \emph{Visibility in degraded visual environments (dve)}, Proc. SPIE Int. Soc. Opt. Eng.
{\bf10642}, 106420S (2018).

\bibitem{Caris17}
M. Caris, W. Johannes, S. Sieger, V. Port, and S. Stanko, \emph{Detection of small UAS with W-band radar}, in Proceedings of the 18th International Radar Symposium , pp. 1–6 (IEEE, New York, 2017).

\bibitem{Trees07}
H. L. Van Trees and K. L. Bell, \emph{Bayesian Bounds for Parameter Estimation and Nonlinear
Filtering/Tracking}, Wiley-IEEE, Piscataway, New York, (2007).

\bibitem{Ziv69}
J. Ziv and M. Zakai, \emph{Some lower bounds on signal parameter estimation}, IEEE Trans. Inform. Theor. {\bf15}, 386 (1969).

\bibitem{Seidman70} L. P. Seidman, \emph{Performance limitations and error calculations for parameter estimation}, Proc. IEEE {\bf58}, 644 (1970).

\bibitem{Chazan75}
D. Chazan, M. Zakai, and J. Ziv, \emph{Improved Lower Bounds on Signal Parameter Estimation}, IEEE Trans. Inform.
Theor. {\bf21}, 90 (1975).

\bibitem{Bellini74}
S. Bellini and G. Tartara, \emph{Bounds on Error in Signal Parameter Estimation}, IEEE Trans. Commun. {\bf22}, 340 (1974).

\bibitem{Weinstein88}
E. Weinstein, \emph{Relations between Belini-Tartara, Chazan-Zakai-Ziv, and Wax-Ziv lower bounds}, IEEE Trans. Inform. Theor. {\bf34}, 342 (1988).

\bibitem{Banchi15}
L. Banchi, S. L. Braunstein, and S. Pirandola, \emph{Quantum Fidelity for Arbitrary Gaussian States}, Phys. Rev. Lett. {\bf 115}, 260501 (2015).

\bibitem{Pirandola08} 
S. Pirandola and S. Lloyd, \emph{Computable bounds for the discrimination of Gaussian states}, Phys. Rev. A {\bf78}, 012331 (2008).

\bibitem{Pereira21}
J. L. Pereira, L. Banchi, S. Pirandola, \emph{Symplectic decomposition from submatrix determinants}, Proc. R. Soc. A. {\bf 477}, 20210513 (2021).


\bibitem{Grace21}
M. R. Grace and Saikat Guha, \emph{Perturbation theory for quantum information}, arXiv:2106.05533 (2021).


\bibitem{Tsang16}
M. Tsang, R. Nair, and X.-M. Lu, \emph{Quantum Theory of Superresolution for Two Incoherent Optical Point Sources}, Phys. Rev. Applied {\bf 6}, 031033 (2016).

\bibitem{Shapiro05}
J. Shapiro, S. Guha, and B. Erkmen, \emph{Ultimate channel capacity of free-space optical communications}, J. Opt. Netw. {\bf4}, 501 (2005).


\end{thebibliography}

\end{document}